\documentclass[pra,twocolumn,nofootinbib,floatfix]{revtex4} 
\usepackage{amsmath}
\usepackage{amssymb}
\usepackage{float}
\usepackage{graphicx}
\usepackage{xspace}

\begin{document}

\title{Photoassociative creation of ultracold heteronuclear $^6$Li$^{40}$K$^*$ molecules}


\author{Armin Ridinger$^1$\footnote{Electronic address: ridinger@ens.fr}}
\author{Saptarishi Chaudhuri$^1$\footnote{Current address: LENS and Dipartimento di Fisica, Universit\`a di Firenze, Via Nello Carrara 1, 50019 Sesto Fiorentino, Italy}}
\author{Thomas Salez$^1$}
\author{Diogo Rio Fernandes$^1$}
\author{Nadia Bouloufa$^2$}
\author{Olivier Dulieu$^2$}
\author{Christophe Salomon$^1$}
\author{Frederic Chevy$^1$}
\affiliation{$^1$Laboratoire Kastler Brossel, \'Ecole Normale Sup\'erieure, CNRS, Universit\'e Pierre et Marie-Curie-Paris 6, 24 rue Lhomond, F-75231 Paris Cedex 05, France}
\affiliation{$^2$Laboratoire Aim\'e Cotton, CNRS, Universit\'e Paris-Sud 11, F-91405 Orsay, France}
\date{\today}

\begin{abstract}
We investigate the formation of weakly bound, electronically excited, heteronuclear $^6$Li$^{40}$K$^*$ molecules by single-photon photoassociation in a magneto-optical trap. We performed trap loss spectroscopy within a range of $325$\,GHz below the Li($2S_{1/2}$)+K($4P_{3/2}$) and Li($2S_{1/2}$)+K($4P_{1/2}$) asymptotic states and observed more than 60 resonances, which we identify as rovibrational levels of 7 of 8 attractive long-range molecular potentials. The long-range dispersion coefficients and rotational constants are derived. We find large molecule formation rates of up to $\sim3.5\times10^{7}$s$^{-1}$, which are shown to be comparable to those for homonuclear $^{40}$K$_2^*$. Using a theoretical model we infer decay rates to the deeply bound electronic ground-state vibrational level $X^1\Sigma^+(v'=3)$ of $\sim5\times10^{4}$s$^{-1}$. Our results pave the way for the production of ultracold bosonic ground-state $^6$Li$^{40}$K molecules which exhibit a large intrinsic permanent electric dipole moment.
\end{abstract}
\maketitle

\section{Introduction}
The recent realization of gases of ultracold polar molecules in their rovibrational ground state~\cite{NiOsp08,DeiGro08} has opened a new frontier in atomic and molecular physics~\cite{CarDeM09,DulGab09}. Due to their long-range anisotropic dipole-dipole interactions and the possibility to trap and manipulate these molecules with external electric fields, they offer fascinating prospects for the realization of new forms of quantum matter~\cite{DamSan03,YiYou00}. Applications to quantum information processing~\cite{DeM02,RabDeM06}, precision measurements~\cite{HudLew06,SheBut08} and ultracold chemistry~\cite{CarDeM09} have been proposed.

The heteronuclear alkali dimer LiK is an excellent candidate for these studies. It has a large dipole moment of 3.6\,D~\cite{AymDul05} in its singlet rovibrational ground state and both of its constituents, Li and K, possess stable fermionic and bosonic isotopes with which dipolar gases of different quantum statistics can be realized. 

While atoms are routinely laser cooled to ultracold temperatures, the complex internal structure of molecules makes this direct method difficult (although possible~\cite{ShuBar10}). So far the most efficient way to produce ultracold molecules has been to associate pre-cooled atoms. Two techniques have been established, namely magnetically tunable Feshbach resonances and photoassociation. Feshbach resonances allow the production of vibrationally excited molecules in the electronic ground state. In this way, ultracold heteronuclear $^6$Li$^{40}$K molecules could recently be produced~\cite{VoiTag09,WilSpi08}. A combination of Feshbach resonances with a multi-photon state transfer may give access to the collisionally stable rovibrational ground state~\cite{DanHal08,NiOsp08}. Photoassociation can directly give access to this state either via single-photon photoassociation and subsequent spontaneous decay~\cite{DeiGro08} or by multi-color photoassociation~\cite{SagJer05}.

In this letter we report on the production of ultracold heteronuclear excited $^6$Li$^{40}$K$^*$ molecules by single-photon photoassociation (PA) in a dual-species magneto-optical trap (MOT). We detect the molecule creation by a loss in the number of trapped atoms, which results from the molecules' spontaneous decay into either a pair of free untrapped atoms or a bound ground-state molecule.

Heteronuclear PA has so far been demonstrated for RbCs$^*$~\cite{KerSag04}, KRb$^*$~\cite{WanQi04}, NaCs$^*$~\cite{HaiKle04}, LiCs$^*$~\cite{DeiGro08} and YbRb$^*$~\cite{NemBau09}. As compared to homonuclear molecules, the PA rate for heteronuclear molecules is typically smaller due to the different range of the excited-state potentials. Whereas two identical atoms in their first excited state interact via the resonant dipole interaction at long range (with potential $V(R)\propto -C_3/R^{3}$), two atoms of different species interact via the van der Waals interaction ($V(R)\propto -C_6/R^{6}$), leading, for the heteronuclear case, to molecule formation at much shorter distances where fewer atom pairs are available. Besides, it has been argued that among the heteronuclear dimers, LiK$^*$ would be particularly difficult to photoassociate due to its small reduced mass and $C_6$ coefficients, which should lead to small PA rates of \textit{e.g.}~two orders of magnitude less than for the heavier dimers RbCs$^*$ and KRb$^*$~\cite{WanStw98}. However, the PA rates we observe in our experiment are similar to those of the comparable experiment with RbCs$^*$~\cite{KerSag04} and those found for homonuclear K$_2^*$. Our theoretical calculations are able to predict the large rates observed.

We perform PA spectroscopy in order to determine the long-range part of the excited-state molecular potentials. Previously, several molecular potentials of LiK have been determined by molecular~\cite{RouAll99,JasKow01,SalRos07} and Feshbach resonance spectroscopy~\cite{WilSpi08,TieKno09}. Our measurements give access to previously undetermined spectroscopic data of high precision and will allow the derivation of more precise molecular potential curves facilitating the search for efficient pathways to produce LiK molecules in the rovibrational ground state.

\begin{figure}[h!]
\centering
\includegraphics[width=1\linewidth]{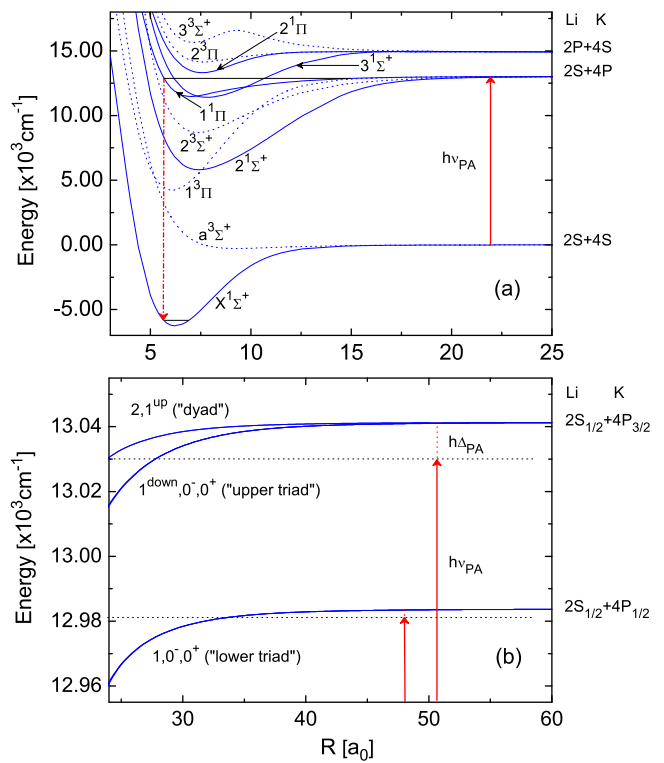}
\caption{(a) Molecular potentials of the LiK molecule for short interatomic separations $R$ ($a_0=0.0529177$\,nm). The upward arrow represents the energy delivered by the PA laser and the downward the spontaneous decay to electronic ground-state molecules. The vibrational state $X^1\Sigma^+(v'=3)$ shown in the figure has a favorable overlap with the addressed excited states due to spatially coincident classical inner turning points. (b) Detailed view of the excited-state potentials, labeled by their Hund's case $c$ quantum numbers $\Omega^{\sigma}$ and an additional classification (``up/down'') for unambiguous distinction. At short range, each of these potentials approaches one of those shown in fig.~(a) as illustrated in ref.~\cite{RomKre09}.}\label{PALIKFIG1}\vspace{-0.25cm}
\end{figure}

Figure~\ref{PALIKFIG1}(a) shows the molecular potentials dissociating to the three lowest electronic asymptotes $2S$+$4S$, $2S$+$4P$ and $2P$+$4S$ of the LiK molecule. They have been calculated as described in ref.~\cite{AymDul05} and connected to the asymptotic form given in ref.~\cite{BusAch87} at large distances ($R>40\,a_0$). Note that as usual in alkali dimers, a strong spin-orbit coupling is expected between the $1^3\Pi$ and the $2^1\Sigma^+$ states due to the crossing of their potential curves around 7.5\,$a_0$. Relevant for our experiment are the eight Hund's case $c$ potential curves dissociating to the $2S_{1/2}$+$4P_{1/2,3/2}$ asymptotes. Figure~\ref{PALIKFIG1}(b) displays their long-range part, which is obtained by diagonalizing the atomic spin-orbit operator in the subspace restricted to the Hund's case $a$ states correlated to $2S$+$4P$, for each of the symmetries $\Omega^{\sigma}=0^+,0^-,1,2$ (where $\Omega$ denotes the quantum number of the projection of the total electronic angular momentum on the molecular axis and $\sigma$ the parity of the electronic wave function through a symmetry with respect to a plane containing the molecular axis). These potentials are all attractive at long range, whereas the curves which dissociate to the asymptotes $2P_{1/2,3/2}$+$4S_{1/2}$ are all repulsive~\cite{WanStw98}. For the relevant asymptotes the dispersion coefficients $C_6$ assume only three different values due to the small atomic fine structure of the Li atom~\cite{MovBeu85}. They have been calculated theoretically~\cite{MovBeu85,BusAch87} and they are determined experimentally in this work. \vspace{-0.25cm}

%
%

\section{Experimental Setup}

The $^6$Li$^{40}$K$^*$ molecules are created by a PA beam which is superimposed with the atoms trapped in the dual-species MOT. The MOT is continuously loaded from a Zeeman slower for $^6$Li and a 2D-MOT for $^{40}$K, as described in ref.~\cite{RidCha11}. We record PA spectra by scanning the PA beam in frequency, simultaneously recording the steady-state atom number of each species via the emitted trap fluorescence. The signature of $^6$Li$^{40}$K$^*$ formation is a decrease of both the $^6$Li and the $^{40}$K fluorescence. The PA laser is scanned red detuned with respect to one of the atomic transitions of $^{40}$K (see fig.~\ref{PALIKFIG1}(b)) and has no effect on a single-species $^6$Li-MOT. The $^6$Li fluorescence signal thus represents a pure heteronuclear PA spectrum, whereas the $^{40}$K fluorescence signal represents the sum of a heteronuclear ($^6$Li$^{40}$K$^*$) and homonuclear ($^{40}$K$_2^*$) PA spectrum. The frequency of the PA laser is recorded by a wavelength meter (High Finesse,~ref.~WS-6) with an absolute accuracy of $\pm250$\,MHz. Additionally, a Fabry-Perot interferometer is used to verify the laser's single-mode operation. 

The PA light is derived from a homemade diode laser-tapered amplifier system. It has a wavelength of 767\,nm and a power of 660\,mW at the output of a single-mode polarization-maintaining fiber. It is collimated and passes four times through the center of the MOT with a total peak intensity of $\sim100$\,W/cm$^2$. The beam diameter of 2.2\,mm (1/e$^2$) was chosen to match the size of the $^6$Li-MOT. Using the feed-forward technique~\cite{FueWal08}, the laser's mode hop free continuous tuning range extends over $\sim35$\,GHz.

For optimum experimental conditions, the PA-induced trap loss needs to be maximized and all other intrinsic losses that compete with it minimized~\cite{RomKre09}. Besides, the frequency of the PA beam needs to be scanned slowly enough ($\sim15$\,MHz/s) to allow the trap loss to reach a quasi-steady state. To achieve these conditions the $^6$Li-MOT is reduced to a small atom number and volume (by lowering the loading rate) and placed at the center of the larger $^{40}$K-MOT. Further, light-induced cold collisions are reduced by using small intensities for the MOT cooling and repumping light ($I_\textrm{\scriptsize cool}^\textrm{\scriptsize Li}\sim1.5I_\textrm{\scriptsize sat}^\textrm{\scriptsize Li}$, $I_\textrm{\scriptsize rep}^\textrm{\scriptsize Li}\sim0.5I_\textrm{\scriptsize sat}^\textrm{\scriptsize Li}$, $I_\textrm{\scriptsize cool}^\textrm{\scriptsize K}\sim10I_\textrm{\scriptsize sat}^\textrm{\scriptsize K}$, $I_\textrm{\scriptsize rep}^\textrm{\scriptsize K}\sim3I_\textrm{\scriptsize sat}^\textrm{\scriptsize K}$ per beam, respectively). The detunings are $\Delta\nu_\textrm{\scriptsize cool}^\textrm{\scriptsize Li}=\Delta\nu_\textrm{\scriptsize rep}^\textrm{\scriptsize Li}\sim-3\Gamma$, $\Delta\nu_\textrm{\scriptsize cool}^\textrm{\scriptsize K}=\Delta\nu_\textrm{\scriptsize rep}^\textrm{\scriptsize K}\sim-4\Gamma$ and the axial magnetic field gradient is 20\,G/cm. These parameters result in $N_\textrm{\scriptsize Li}\sim 5\times10^8$ and $N_\textrm{\scriptsize K}\sim 2.5\times10^9$ trapped atoms with central atomic densities of $n_\textrm{\scriptsize Li}\sim 7\times10^{10}$\,cm$^{-3}$ and $n_\textrm{\scriptsize K}\sim 5\times10^{10}$\,cm$^{-3}$ and temperatures of $T_\textrm{\scriptsize Li}\sim1.2$\,mK and $T_\textrm{\scriptsize K}\sim300\,\mu$K, respectively. At these temperatures only heteronuclear collisions of $s$- and $p$-wave character (\textit{i.e.}, $\ell=0,1$, where $\ell$ is the rotational angular momentum of the atom pair) reach sufficiently short internuclear distances to allow for PA (the height of the $d$-wave rotational barrier being 13.4\,mK). If $J$ is the total angular momentum quantum number of the atom pair (including electronic angular momentum and rotation), molecule formation is thus restricted to the rotational levels $J=0,1,2$ for electronic states with $\Omega=0$, $J=1,2,3$ for $\Omega=1$ and $J=2,3$ for $\Omega=2$. \vspace{-0.5cm}

\section{Results}

\begin{figure*}[t]
\begin{center}
\includegraphics[width=2\columnwidth]{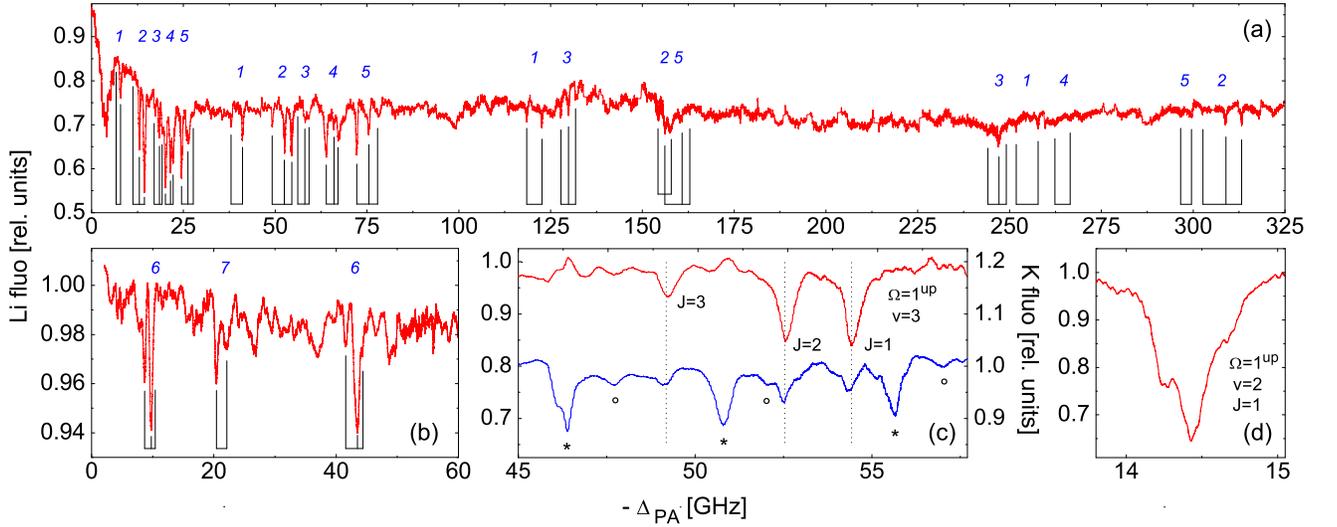}\vspace{-0.5cm}
\end{center}
\caption{Heteronuclear PA trap loss spectra of $^6$Li$^{40}$K$^*$ below the asymptotes $2S_{1/2}$+$4P_{3/2}$ (a) and $2S_{1/2}$+$4P_{1/2}$ (b). The spectra contain seven vibrational series (labeled $N=1,...,7$) with resolved rotational structure, whose assignment is given in table~\ref{Resonances}. (c) Zoom on the heteronuclear (upper trace, left axis) and heteronuclear+homonuclear (lower trace, right axis) PA spectrum below the $2S_{1/2}$+$4P_{3/2}$ asymptote showing the rotational structure of the $\Omega=1^\textrm{\tiny up},v=3$ vibrational state of $^6$Li$^{40}$K$^*$ ($v$ denoting the vibrational quantum number counted from dissociation) and three vibrational $0^+_\textrm{\scriptsize u}$ levels of $^{40}$K$_2^*$, which show a resolved hyperfine ($*$ and $\circ$) but no rotational structure. (d) Zoom on the $\Omega=1^\textrm{\tiny up},v=2,J=1$ resonance of $^6$Li$^{40}$K$^*$, showing a nearly resolved hyperfine structure. The PA detuning $\Delta_\textrm{\scriptsize PA}$ is specified relative to the $^{40}$K atomic transitions $4S_{1/2}(F=9/2)\rightarrow 4P_{3/2}(F'=11/2)$ (a,c,d) and $4S_{1/2}(F=9/2)\rightarrow 4P_{1/2}(F'=9/2)$ (b).} \label{PALIKFIG2}
\end{figure*}

Figure~\ref{PALIKFIG2}~(a-d) shows a compilation of our recorded spectroscopic data. Figures 2~(a) and (b) depict the heteronuclear PA spectra near the dissociation limits $2S_{1/2}$+$4P_{3/2}$ and $2S_{1/2}$+$4P_{1/2}$ for PA detunings $\Delta_\textrm{\scriptsize PA}$ between $0$ and $-325$\,GHz and between $0$ and $-60$\,GHz, respectively. The graphs represent, respectively, an average of $\sim6$ and $\sim20$ recorded spectra for noise reduction and have been recorded in pieces and stitched together. The spectra contain $68$ resonances whose contrasts decrease and whose mutual separations increase with increasing detuning. The maximum contrast amounts to $\sim35\%$ and is obtained for a detuning of $\Delta_\textrm{\scriptsize PA}=-14.4\,$GHz (see Fig.~\ref{PALIKFIG2}~(d)). The observed resonance widths (FWHM) vary between 80 and 300\,MHz, primarily due to unresolved molecular hyperfine structure.

%
%

We have also recorded the heteronuclear+homonuclear PA spectra appearing on the K fluorescence signal, which contain all the resonances of fig.~\ref{PALIKFIG2} (a,b) as well. A comparison between the two spectra is shown for a small part in fig.~\ref{PALIKFIG2} (c). This figure shows comparable contrasts for the heteronuclear $^6$Li$^{40}$K$^*$ and homonuclear $^{40}$K$_2^*$ PA signals. We identify the observed $^{40}$K$_2^*$ resonances as excitations to $0^+_\textrm{\scriptsize u}$ states~\cite{Rid11}. 

In the heteronuclear spectra of fig.~\ref{PALIKFIG2} (a,b) we identify seven vibrational series (labeled with numbers), corresponding to seven of the eight molecular potentials dissociating to the $2S_{1/2}$+$4P_{1/2,3/2}$ asymptotes (see fig.~\ref{PALIKFIG1} (b)). Each series contains up to five resonances, which appear in doublets or triplets due to resolved rotational structure. This structure is shown more clearly in fig.~\ref{PALIKFIG2} (c) for a particular vibrational state. Some of the observed rovibrational resonances have a further substructure resulting from hyperfine interactions, which is shown for a particular resonance in fig.~\ref{PALIKFIG2} (d).

In fig.~\ref{PALIKFIG2} (a,b) and table~\ref{Resonances} we present an assignment of the observed resonances, which was obtained by the combination of different assignment rules: 
the first is the rotational progression law $E_\textrm{\scriptsize rot}=B_{v}[J(J+1)-\Omega^2]$, with $J=\Omega,\Omega+1,...$ for Hund's case $c$ molecules~\cite{RomKre09} combined with our theoretical calculations of the rotational constants $B_{v}$. It allowed us to identify rotational progressions and to assign some $J$ and $\Omega$ based on the rotational spacing. The identification of the $\Omega=2$ vibrational series (series $1$ in fig.~\ref{PALIKFIG2} (a)) is particularly easy, because only two rotational lines per vibrational level are expected, as opposed to three for all other series. The second is the semi-classical LeRoy-Bernstein (LRB) law~\cite{LeRBer70,Stw70} (see Eq.~(\ref{LRB})) combined with the available calculated $C_6$ coefficients~\cite{MovBeu85,BusAch87}. It allowed us to identify vibrational progressions and to assign some $v$ and $\Omega$ based on the vibrational spacing. The third is the hyperfine structure law $E_\textrm{\scriptsize hfs}\propto \Omega/[J(J+1)]$ for $\Omega=1$ and $E_\textrm{\scriptsize hfs}\approx0$ for $\Omega=0$~\cite{WilTie96,GroPas09}. It predicts small widths for resonances with $\Omega=0$ and particularly large widths for those with $\Omega=1,J=1$ making their identification possible. The fourth is the expected similar contrast pattern of the rotational lines of the same vibrational series, which helped us to identify vibrational progressions. 

%
\begin{table}[h]
		\begin{tabular}{ccccrrp{0.01cm}||p{0.01cm}ccccrr} \hline\hline
    $N$ & $\Omega$ & $v$ & $J$ & $-\Delta_\textrm{\scriptsize PA}\,$ & Contr. & & & $N$ & $\Omega$ & $v$ & $J$ & $-\Delta_\textrm{\scriptsize PA}\,$ & Contr.\\
		   &      &     &     & [GHz]                              & [\%] \ \,     & & &        &  &     &     & [GHz]                              & [\%] \ \, \\ \hline
$1$ &  $2$   &1&2  &	$0.37$	  &  1.1 & & & $4$ &  $0^-$     &2&2  &	$3.03$	  & 21.4\\
    &        &2&3  &  $6.60$	  &  2.3 & & &     &            & &1  &	$3.66$	  & 22.7\\
    &        & &2  &	$7.88$	  &  9.4 & & &     &            &3&2  &	$20.11$   & 25.0\\
    &        &3&3  &	$38.00$	  &  5.6 & & &     &            & &1  &	$21.48$   & 20.4\\
    &        & &2  &	$41.08$	  & 10.6 & & &     &            & &0  &	$22.16$   & 17.4\\
    &        &4&3  &	$118.49$	&  6.3 & & &     &            &4&2  &	$63.99$   & 14.7\\
    &        & &2  &	$122.67$	&  6.3 & & &     &            & &1  &	$66.06$   & 5.0\\
    &        &5&3  &	$251.80$	&  3.7 & & &     &            & &0  &	$67.13$   & 7.6\\
    &        & &2  &	$257.81$	&  4.3 & & &     &            &6&2  &	$262.37$  & 1.9\\\cline{1-7}
$2$ & $1^\textrm{\tiny up}$ &1&2  &	$0.37$	  &  1.1 & & &     &        & &0  &	$266.75$  & 1.9\\\cline{8-14}
    &        & &1  &	$1.34$	  &  6.1 & & & $5$ & $1^\textrm{\tiny down}$  &2&3  &	$4.12$	  & 23.9\\
    &        &2&3  &	$11.27$   &  2.4 & & &     &            & &2  &	$4.85$	  & 14.9\\
    &        & &2  &	$13.01$   & 18.5 & & &     &            & &1  &	$5.90$	  &  4.6\\
    &        & &1  &	$14.40$   & 35.0 & & &     &            &3&3  &	$24.52$   & 19.8\\
    &        &3&3  &	$49.20$   &  6.5 & & &     &            & &2  &	$26.15$   &  9.8\\
    &        & &2  &	$52.47$   & 14.1 & & &     &            & &1  &	$27.72$   &  3.0\\
    &        & &1  &	$54.53$   & 14.8 & & &     &            &4&3  &	$72.26$   & 15.4\\
    &        &4&2  &	$154.35$  &  5.4 & & &     &            & &2  &	$75.55$   & 10.0\\
    &        & &1  &	$157.40$  &  4.7 & & &     &            & &1  &	$77.87$   &  5.0\\
    &        &5&3  &	$302.50$  &  2.0 & & &     &            &5&3  &	$156.12$  &  6.1\\
    &        & &2  &	$308.75$  &  3.9 & & &     &            & &2  &	$160.20$  &  1.4\\
    &        & &1  &	$313.26$  &  4.6 & & &     &            & &1  &	$162.91$  &  1.9\\\cline{1-7}
$3$ & $0^+$  &2&2  &	$1.34$	  &  6.1 & & &     &            &6&2  &	$296.51$  &  1.8\\
    &        & &1  &	$2.35$	  & 11.0 & & &     &            & &1  &	$299.57$  &  1.8\\\cline{8-14}
    &        & &0  &	$3.03$	  & 21.4 & & & $6$ & $0^+$      &2&2  &  $8.66$   &  3.9\\
    &        &3&2  &	$17.23$   &  6.2 & & &     &            & &1  &  $9.74$   &  5.8\\
    &        & &1  &	$18.51$   & 10.6 & & &     &            & &0  &  $10.37$  &  0.8\\
    &        & &0  &	$19.25$   &  8.4 & & &     &            &3&2  &	$41.56$   &  1.6\\
    &        &4&2  &	$56.30$   &  2.0 & & &     &            & &1  &	$43.40$   &  5.1\\
    &        & &1  &	$58.17$   &  6.6 & & &     &            & &0  &	$44.31$   &  2.0\\\cline{8-14}
    &        & &0  &	$59.21$   &  5.6 & & & $7$ & $1$        &2&2  &	$20.42$	  &  3.6\\
    &        &5&2  &	$127.76$  &  4.0 & & &     &            & &1  &	$21.93$   &  2.0\\\cline{8-14}
    &        & &1  &	$129.91$  &  7.8 & & &     &            & &   &           &     \\
    &        & &0  &	$131.80$  &  5.0 & & &     &            & &   &           &     \\
    &        &6&2  &	$244.13$  &  2.0 & & &     &            & &   & \multicolumn{2}{c}{Accuracy}\\
    &        & &1  &	$247.07$  &  5.2 & & &     &            & &   & $\pm 0.25$&  $\pm1.0$\\
    &        & &0  &	$249.01$  &  2.2 & & &     &            & &   &           &     \\\hline\hline         
  \end{tabular}
	\caption{PA resonances of $^6$Li$^{40}$K$^*$ observed below the $2S_{1/2}$+$4P_{1/2,3/2}$ asymptotes and their contrasts. $N$ denotes the number of the vibrational series given in fig.~\ref{PALIKFIG2} (a,b).}
	\label{Resonances}\vspace{-0.25cm}
\end{table}
%

An application of the assignment rules allowed us to identify the observed vibrational series and to assign their quantum numbers except the parity $\sigma$ of the $\Omega=0$ electronic states. $\sigma$ can be determined from an analysis of the relative strength of the rotational lines: due to the selection rules, the parity of the total wave function of the system, \textit{i.e.}~the product of $\sigma$ and $(-)^\ell$ for the rotational part, changes sign during the transition. Further, $\sigma$ is conserved, namely only $X^1\Sigma^+ (0^+) \rightarrow 0^+$ and $a^3\Sigma^+(0^-) \rightarrow 0^-$ are allowed for parallel transitions. In our experiment, $s$-wave collisions dominate, such that the total parity is $+$ ($-$) for the former (the latter) initial state. The parallel transition  $X^1\Sigma^+ (0^+,\ell=0) \rightarrow (0^+,\ell=1)$ is thus allowed enhancing then the $J=1$ line, while the parallel transition $a^3\Sigma^+ (0^-,\ell=0) \rightarrow (0^-,\ell=1)$ is forbidden. Under the same approximation, the perpendicular transition $a^3\Sigma^+(1,\ell=0) \rightarrow (0^-,\ell=0)$ is allowed and enhances the $J=0$ line in the spectrum. Therefore we assign the $\Omega=0$ series with pronounced (reduced) $J=1$ line to the excited $0^+$ (0$^-$) states.\vspace{-0.25cm}

\section{Discussion}

\begin{figure}[h!]
\centering
\includegraphics[width=0.8\linewidth]{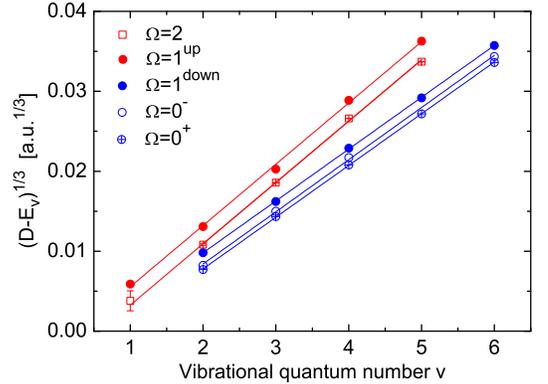}
\caption{Plot of the $1/3$-rd power of the measured binding energies $D-E_v=-(h\Delta_\textrm{\scriptsize PA}-E_\textrm{\scriptsize rot})$ (symbols) as a function of the vibrational quantum number counted from the dissociation limit for the five vibrational series dissociating to the $2S_{1/2}$+$4P_{3/2}$ asymptote. The slopes of the linear fits (solid lines) yield the dispersion coefficients $C_6$ according to eq.~(\ref{LRB}). The nearly identical slopes of the triad and dyad potentials demonstrate the equality of the respective $C_6$ coefficients.}\label{PALIKFIG3}\vspace{-0.5cm}
\end{figure}

Having assigned all observed resonances, the parameters of the different molecular potentials can be derived. We infer the $C_6$ coefficients from the measured vibrational binding energies $D-E_v=-(h\Delta_\textrm{\scriptsize PA}-E_\textrm{\scriptsize rot})$ ($D$ denoting the dissociation energy and $E_v$ the energy of the vibrational level $v$), using the LRB formula~\cite{LeRBer70,Stw70}
\begin{eqnarray}\label{LRB}
	D-E_v=A_6(v_D-v)^{3},
\end{eqnarray}
with $A_6=64\pi^3\hbar^3C_6/\left[B(2/3,1/2)\sqrt{2\mu C_6}\right]^{3}$, where $B$ denotes the Beta-function ($B(2/3,1/2)\approx2.587$), $\mu$ is the reduced mass, and $v_D$ the vibrational quantum number at dissociation (a constant between $0$ and $1$ so that the most weakly bound state has $v=1$). Figure~\ref{PALIKFIG3} shows the plots of the $1/3$-rd power of the binding energies as a function of the vibrational quantum number for the five vibrational series dissociating to the $2S_{1/2}$+$4P_{3/2}$ asymptote. The plots are predicted to follow straight lines whose slopes yield: $C_6=9170\pm940\,$a.u.~and $C_6=9240\pm960\,$a.u.~for the dyad potentials $\Omega=2,1^\textrm{\scriptsize up}$, $C_6=25220\pm600\,$a.u., $C_6=25454\pm720\,$a.u.~and $C_6=24310\pm1710\,$a.u.~for the upper triad potentials $\Omega=1^\textrm{\scriptsize down},0^+,0^-$ and $C_6=12860\pm660\,$a.u.~for the lower triad potential $\Omega=0^+$ (not shown in fig.~\ref{PALIKFIG3}), respectively, where the uncertainties represent statistical uncertainties for the fits. These values are in good agreement with the respective theoretical values $C_6=9800\,$a.u., $C_6=25500\,$a.u.~and $C_6=13830\,$a.u.~predicted by Bussery et al.~\cite{BusAch87}. The agreement with the values $C_6=9520\,$a.u., $C_6=22000\,$a.u.~and $C_6=15420\,$a.u.~predicted by Movre et al.~\cite{MovBeu85} is not as good. The two predictions differ in their treatment of the interaction between the two asymptotes $2S$+$4P$ and $2P$+$4S$, which is taken into account in ref.~\cite{BusAch87} only, hinting its significance. 

%
%
The uncertainty of the derived $C_6$ coefficients results from the following effects. First, the heteronuc
lear nature of LiK and its small $C_6$ coefficients lead to molecule formation at small internuclear separations (of $R_\textrm{\scriptsize eff}=\hbar/\sqrt{2\mu B_{v}}\sim 18\,a_0$ at $\Delta_\textrm{\scriptsize PA}=-300\,$GHz) at which the exchange interaction and higher-order terms in the long-range multipole expansion of the molecular potential become important, which are neglected by the LRB law. Second, the small reduced mass of LiK leads to a low density of vibrational states and thus to a small number of states with long-range character available for fitting.

\begin{figure}[h!]
\centering
\includegraphics[width=0.8\linewidth]{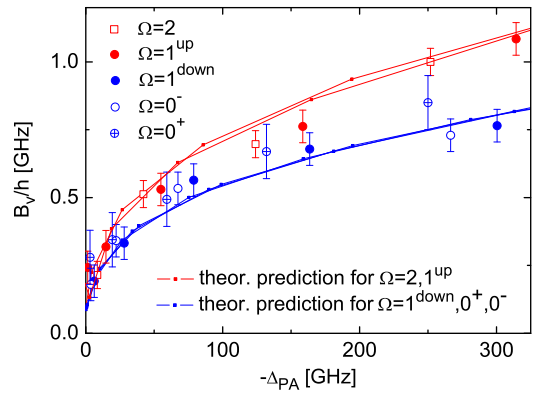}
\caption{Measured rotational constants (symbols) for the observed excited molecular states below the $2S_{1/2}$+$4P_{3/2}$ asymptote and their theoretical predictions for computed vibrational levels (dots, the lines serve to guide the eye), derived from the potential curves of fig.~\ref{PALIKFIG1}.}\label{PALIKFIG4}\vspace{-0.25cm}
\end{figure}

The measured rotational splittings allow us to infer the rotational constants and to confirm the assignments above. They are shown in fig.~\ref{PALIKFIG4} for the five vibrational series below the $2S_{1/2}$+$4P_{3/2}$ asymptote, together with their theoretical predictions, which we have derived from the potential curves of fig.~\ref{PALIKFIG1}. The agreement between the measured and predicted values is reasonable. The error bars account for the imprecision of the wavelength determination and of the resonance positions due to the unresolved hyperfine structure.
Deviations from the theoretical predictions are likely to be due to the multichannel character of the vibrational levels.

%
%

We have determined the $^6$Li$^{40}$K$^*$ molecule formation rate from the steady-state depletion of the $^{6}$Li atom number induced by PA. For the resonance shown in fig.~\ref{PALIKFIG2} (d) we obtain a lower bound of $\beta_\textrm{\scriptsize PA}n_\textrm{\scriptsize K}N_\textrm{\scriptsize Li}^\textrm{\scriptsize PA}\sim3.5\times10^{7}$\,s$^{-1}$ and a PA rate coefficient $\beta_\textrm{\scriptsize PA}=(2.2\pm1.1)\times10^{-12}\,$cm$^3$/s. This coefficient is larger by about a factor of two than the one found in the experiment with RbCs$^*$~\cite{KerSag04}, showing that PA rates for LiK$^*$ are much more favorable than previously expected~\cite{WanStw98}, confirming the trend discussed in ref.~\cite{AziAym04}. Using the approach described in ref.~\cite{PilCru97}
we estimate the total PA rate coefficient for a computed $1^\textrm{\scriptsize up}$ level with $-21$\,GHz detuning to $1.6\times 10^{-12}\,$cm$^3$/s~\cite{Rid11}, in agreement with our measured value. The associated $^6$Li$^{40}$K$^*$ molecule formation rate is also found to be comparable to that for $^{40}$K$_2^*$, which is derived from fig.~\ref{PALIKFIG2} (c) to be $\sim5.3\times10^7\,$s$^{-1}$, despite the much longer range of the excited $^{40}$K$_2^*$ molecular potential. Using our model, which reproduces the observed rovibrational structure, we infer the rates for decay into bound levels of the $X^1\Sigma^+$ state. The rates are found largest for the decay into the most weakly bound level, but are still significant for the decay into deeply bound levels such as the $X^1\Sigma^+(v'=3)$ level ($v'$ counted from the potential bottom) for which it is $5\times10^{4}$\,s$^{-1}$ (see fig.~\ref{PALIKFIG1}). Since the $^6$Li$^{40}$K$^*$ molecule formation rate saturates in our experiment at moderate PA intensities of $\sim80$\,W/cm$^2$, efficient coherent multi-photon population transfers to the molecular rovibrational ground state can be expected. \vspace{-0.1cm}


\section{Conclusion}

In summary, we have investigated single-photon photoassociation of excited heteronuclear $^6$Li$^{40}$K$^*$ molecules. We have recorded photoassociation spectra and assigned all observed resonances. We have derived the long-range dispersion coefficients and rotational constants, which agree with the theoretical predictions of ref.~\cite{BusAch87} and our calculations, respectively. In particular, we have observed large formation rates for the heteronuclear $^6$Li$^{40}$K$^*$ molecules which are comparable to those found for homonuclear $^{40}$K$_2^*$. These rates promise efficient creation of rovibrational ground-state molecules and show that photoassociation is an attractive alternative to Feshbach resonances, since those have a very small width for $^6$Li$^{40}$K and are thus difficult to control~\cite{WilSpi08}.

For future research it will be interesting to combine ours and previously recorded data on LiK and to refine the molecular potentials. Spectroscopic data is available for the potential $1^1\Pi$~\cite{PasJas98}, which correlates with the $\Omega=1^\textrm{\scriptsize up}$ potential, for which we measured the binding energies of the five previously undetermined least-bound vibrational states, such that a complete set of vibrational levels is now available for a high-precision refinement of this potential.\vspace{-0.5cm}

\acknowledgments
The authors thank L.~Pruvost and J.~Deiglmayr for insightful discussions, U.~Eismann for assistance with the experiment and M.~Aymar for providing us with unpublished LiK potential curves and transition dipole moments. Support from ESF Euroquam (FerMix), SCALA, ANR FABIOLA, R\'egion Ile de France (IFRAF), ERC Ferlodim and Institut Universitaire de France is acknowledged. A.~R.~acknowledges funding from the German Federal Ministry of Education and Research and D.~R.~F.~from Funda\c{c}\~{a}o para a Ci\^encia e Tecnologia and Funda\c{c}\~{a}o Calouste Gulbenkian.

%
\bibliography{bibliography}

\begin{thebibliography}{38}
\expandafter\ifx\csname natexlab\endcsname\relax\def\natexlab#1{#1}\fi
\expandafter\ifx\csname bibnamefont\endcsname\relax
  \def\bibnamefont#1{#1}\fi
\expandafter\ifx\csname bibfnamefont\endcsname\relax
  \def\bibfnamefont#1{#1}\fi
\expandafter\ifx\csname citenamefont\endcsname\relax
  \def\citenamefont#1{#1}\fi
\expandafter\ifx\csname url\endcsname\relax
  \def\url#1{\texttt{#1}}\fi
\expandafter\ifx\csname urlprefix\endcsname\relax\def\urlprefix{URL }\fi
\providecommand{\bibinfo}[2]{#2}
\providecommand{\eprint}[2][]{\url{#2}}

\bibitem[{\citenamefont{Ni et~al.}(2008)\citenamefont{Ni, Ospelkaus,
  de~Miranda, Pe'er, Neyenhuis, Zirbel, Kotochigova, Julienne, Jin, and
  Ye}}]{NiOsp08}
\bibinfo{author}{\bibfnamefont{K.-K.} \bibnamefont{Ni}},
  \bibinfo{author}{\bibfnamefont{S.}~\bibnamefont{Ospelkaus}},
  \bibinfo{author}{\bibfnamefont{M.~H.~G.} \bibnamefont{de~Miranda}},
  \bibinfo{author}{\bibfnamefont{A.}~\bibnamefont{Pe'er}},
  \bibinfo{author}{\bibfnamefont{B.}~\bibnamefont{Neyenhuis}},
  \bibinfo{author}{\bibfnamefont{J.~J.} \bibnamefont{Zirbel}},
  \bibinfo{author}{\bibfnamefont{S.}~\bibnamefont{Kotochigova}},
  \bibinfo{author}{\bibfnamefont{P.~S.} \bibnamefont{Julienne}},
  \bibinfo{author}{\bibfnamefont{D.~S.} \bibnamefont{Jin}}, \bibnamefont{and}
  \bibinfo{author}{\bibfnamefont{J.}~\bibnamefont{Ye}},
  \bibinfo{journal}{Science} \textbf{\bibinfo{volume}{322}},
  \bibinfo{pages}{231} (\bibinfo{year}{2008}).

\bibitem[{\citenamefont{Deiglmayr et~al.}(2008)\citenamefont{Deiglmayr,
  Grochola, Repp, M\"ortlbauer, Gl\"uck, Lange, Dulieu, Wester, and
  Weidem\"uller}}]{DeiGro08}
\bibinfo{author}{\bibfnamefont{J.}~\bibnamefont{Deiglmayr}},
  \bibinfo{author}{\bibfnamefont{A.}~\bibnamefont{Grochola}},
  \bibinfo{author}{\bibfnamefont{M.}~\bibnamefont{Repp}},
  \bibinfo{author}{\bibfnamefont{K.}~\bibnamefont{M\"ortlbauer}},
  \bibinfo{author}{\bibfnamefont{C.}~\bibnamefont{Gl\"uck}},
  \bibinfo{author}{\bibfnamefont{J.}~\bibnamefont{Lange}},
  \bibinfo{author}{\bibfnamefont{O.}~\bibnamefont{Dulieu}},
  \bibinfo{author}{\bibfnamefont{R.}~\bibnamefont{Wester}}, \bibnamefont{and}
  \bibinfo{author}{\bibfnamefont{M.}~\bibnamefont{Weidem\"uller}},
  \bibinfo{journal}{Phys. Rev. Lett.} \textbf{\bibinfo{volume}{101}},
  \bibinfo{pages}{133004} (\bibinfo{year}{2008}).

\bibitem[{\citenamefont{Carr et~al.}(2009)\citenamefont{Carr, DeMille, Krems,
  and Ye}}]{CarDeM09}
\bibinfo{author}{\bibfnamefont{L.~D.} \bibnamefont{Carr}},
  \bibinfo{author}{\bibfnamefont{D.}~\bibnamefont{DeMille}},
  \bibinfo{author}{\bibfnamefont{R.~V.} \bibnamefont{Krems}}, \bibnamefont{and}
  \bibinfo{author}{\bibfnamefont{J.}~\bibnamefont{Ye}}, \bibinfo{journal}{New
  J. Phys.} \textbf{\bibinfo{volume}{11}}, \bibinfo{pages}{055049}
  (\bibinfo{year}{2009}).

\bibitem[{\citenamefont{Dulieu and Gabbanini}(2009)}]{DulGab09}
\bibinfo{author}{\bibfnamefont{O.}~\bibnamefont{Dulieu}} \bibnamefont{and}
  \bibinfo{author}{\bibfnamefont{C.}~\bibnamefont{Gabbanini}},
  \bibinfo{journal}{Rep. Prog. Phys.} \textbf{\bibinfo{volume}{72}},
  \bibinfo{pages}{086401} (\bibinfo{year}{2009}).

\bibitem[{\citenamefont{Damski et~al.}(2003)\citenamefont{Damski, Santos,
  Tiemann, Lewenstein, Kotochigova, Julienne, and Zoller}}]{DamSan03}
\bibinfo{author}{\bibfnamefont{B.}~\bibnamefont{Damski}},
  \bibinfo{author}{\bibfnamefont{L.}~\bibnamefont{Santos}},
  \bibinfo{author}{\bibfnamefont{E.}~\bibnamefont{Tiemann}},
  \bibinfo{author}{\bibfnamefont{M.}~\bibnamefont{Lewenstein}},
  \bibinfo{author}{\bibfnamefont{S.}~\bibnamefont{Kotochigova}},
  \bibinfo{author}{\bibfnamefont{P.}~\bibnamefont{Julienne}}, \bibnamefont{and}
  \bibinfo{author}{\bibfnamefont{P.}~\bibnamefont{Zoller}},
  \bibinfo{journal}{Phys. Rev. Lett.} \textbf{\bibinfo{volume}{90}},
  \bibinfo{pages}{110401} (\bibinfo{year}{2003}).

\bibitem[{\citenamefont{Yi and You}(2000)}]{YiYou00}
\bibinfo{author}{\bibfnamefont{S.}~\bibnamefont{Yi}} \bibnamefont{and}
  \bibinfo{author}{\bibfnamefont{L.}~\bibnamefont{You}},
  \bibinfo{journal}{Phys. Rev. A} \textbf{\bibinfo{volume}{61}},
  \bibinfo{pages}{041604} (\bibinfo{year}{2000}).

\bibitem[{\citenamefont{DeMille}(2002)}]{DeM02}
\bibinfo{author}{\bibfnamefont{D.}~\bibnamefont{DeMille}},
  \bibinfo{journal}{Phys. Rev. Lett.} \textbf{\bibinfo{volume}{88}},
  \bibinfo{pages}{067901} (\bibinfo{year}{2002}).

\bibitem[{\citenamefont{Rabl et~al.}(2006)\citenamefont{Rabl, DeMille, Doyle,
  Lukin, Schoelkopf, and Zoller}}]{RabDeM06}
\bibinfo{author}{\bibfnamefont{P.}~\bibnamefont{Rabl}},
  \bibinfo{author}{\bibfnamefont{D.}~\bibnamefont{DeMille}},
  \bibinfo{author}{\bibfnamefont{J.~M.} \bibnamefont{Doyle}},
  \bibinfo{author}{\bibfnamefont{M.~D.} \bibnamefont{Lukin}},
  \bibinfo{author}{\bibfnamefont{R.~J.} \bibnamefont{Schoelkopf}},
  \bibnamefont{and} \bibinfo{author}{\bibfnamefont{P.}~\bibnamefont{Zoller}},
  \bibinfo{journal}{Phys. Rev. Lett.} \textbf{\bibinfo{volume}{97}},
  \bibinfo{pages}{033003} (\bibinfo{year}{2006}).

\bibitem[{\citenamefont{Hudson et~al.}(2006)\citenamefont{Hudson, Lewandowski,
  Sawyer, and Ye}}]{HudLew06}
\bibinfo{author}{\bibfnamefont{E.~R.} \bibnamefont{Hudson}},
  \bibinfo{author}{\bibfnamefont{H.~J.} \bibnamefont{Lewandowski}},
  \bibinfo{author}{\bibfnamefont{B.~C.} \bibnamefont{Sawyer}},
  \bibnamefont{and} \bibinfo{author}{\bibfnamefont{J.}~\bibnamefont{Ye}},
  \bibinfo{journal}{Phys. Rev. Lett.} \textbf{\bibinfo{volume}{96}},
  \bibinfo{pages}{143004} (\bibinfo{year}{2006}).

\bibitem[{\citenamefont{Shelkovnikov et~al.}(2008)\citenamefont{Shelkovnikov,
  Butcher, Chardonnet, and Amy-Klein}}]{SheBut08}
\bibinfo{author}{\bibfnamefont{A.}~\bibnamefont{Shelkovnikov}},
  \bibinfo{author}{\bibfnamefont{R.~J.} \bibnamefont{Butcher}},
  \bibinfo{author}{\bibfnamefont{C.}~\bibnamefont{Chardonnet}},
  \bibnamefont{and}
  \bibinfo{author}{\bibfnamefont{A.}~\bibnamefont{Amy-Klein}},
  \bibinfo{journal}{Phys. Rev. Lett.} \textbf{\bibinfo{volume}{100}},
  \bibinfo{pages}{150801} (\bibinfo{year}{2008}).

\bibitem[{\citenamefont{Aymar and Dulieu}(2005)}]{AymDul05}
\bibinfo{author}{\bibfnamefont{M.}~\bibnamefont{Aymar}} \bibnamefont{and}
  \bibinfo{author}{\bibfnamefont{O.}~\bibnamefont{Dulieu}},
  \bibinfo{journal}{J. Chem. Phys.} \textbf{\bibinfo{volume}{122}},
  \bibinfo{pages}{204302} (\bibinfo{year}{2005}).

\bibitem[{\citenamefont{Shuman et~al.}(2010)\citenamefont{Shuman, Barry, and
  DeMille}}]{ShuBar10}
\bibinfo{author}{\bibfnamefont{E.~S.} \bibnamefont{Shuman}},
  \bibinfo{author}{\bibfnamefont{J.~F.} \bibnamefont{Barry}}, \bibnamefont{and}
  \bibinfo{author}{\bibfnamefont{D.}~\bibnamefont{DeMille}},
  \bibinfo{journal}{Nature} \textbf{\bibinfo{volume}{467}},
  \bibinfo{pages}{820} (\bibinfo{year}{2010}).

\bibitem[{\citenamefont{Voigt et~al.}(2009)\citenamefont{Voigt, Taglieber,
  Costa, Aoki, Wieser, H\"ansch, and Dieckmann}}]{VoiTag09}
\bibinfo{author}{\bibfnamefont{A.-C.} \bibnamefont{Voigt}},
  \bibinfo{author}{\bibfnamefont{M.}~\bibnamefont{Taglieber}},
  \bibinfo{author}{\bibfnamefont{L.}~\bibnamefont{Costa}},
  \bibinfo{author}{\bibfnamefont{T.}~\bibnamefont{Aoki}},
  \bibinfo{author}{\bibfnamefont{W.}~\bibnamefont{Wieser}},
  \bibinfo{author}{\bibfnamefont{T.~W.} \bibnamefont{H\"ansch}},
  \bibnamefont{and}
  \bibinfo{author}{\bibfnamefont{K.}~\bibnamefont{Dieckmann}},
  \bibinfo{journal}{Phys. Rev. Lett.} \textbf{\bibinfo{volume}{102}},
  \bibinfo{pages}{020405} (\bibinfo{year}{2009}).

\bibitem[{\citenamefont{Wille et~al.}(2008)\citenamefont{Wille, Spiegelhalder,
  Kerner, Naik, Trenkwalder, Hendl, Schreck, Grimm, Tiecke, Walraven
  et~al.}}]{WilSpi08}
\bibinfo{author}{\bibfnamefont{E.}~\bibnamefont{Wille}},
  \bibinfo{author}{\bibfnamefont{F.~M.} \bibnamefont{Spiegelhalder}},
  \bibinfo{author}{\bibfnamefont{G.}~\bibnamefont{Kerner}},
  \bibinfo{author}{\bibfnamefont{D.}~\bibnamefont{Naik}},
  \bibinfo{author}{\bibfnamefont{A.}~\bibnamefont{Trenkwalder}},
  \bibinfo{author}{\bibfnamefont{G.}~\bibnamefont{Hendl}},
  \bibinfo{author}{\bibfnamefont{F.}~\bibnamefont{Schreck}},
  \bibinfo{author}{\bibfnamefont{R.}~\bibnamefont{Grimm}},
  \bibinfo{author}{\bibfnamefont{T.~G.} \bibnamefont{Tiecke}},
  \bibinfo{author}{\bibfnamefont{J.~T.~M.} \bibnamefont{Walraven}},
  \bibnamefont{et~al.}, \bibinfo{journal}{Phys. Rev. Lett.}
  \textbf{\bibinfo{volume}{100}}, \bibinfo{pages}{053201}
  (\bibinfo{year}{2008}).

\bibitem[{\citenamefont{Danzl et~al.}(2008)\citenamefont{Danzl, Haller,
  Gustavsson, Mark, Hart, Bouloufa, Dulieu, Ritsch, and N\"agerl}}]{DanHal08}
\bibinfo{author}{\bibfnamefont{J.}~\bibnamefont{Danzl}},
  \bibinfo{author}{\bibfnamefont{E.}~\bibnamefont{Haller}},
  \bibinfo{author}{\bibfnamefont{M.}~\bibnamefont{Gustavsson}},
  \bibinfo{author}{\bibfnamefont{M.~J.} \bibnamefont{Mark}},
  \bibinfo{author}{\bibfnamefont{R.}~\bibnamefont{Hart}},
  \bibinfo{author}{\bibfnamefont{N.}~\bibnamefont{Bouloufa}},
  \bibinfo{author}{\bibfnamefont{O.}~\bibnamefont{Dulieu}},
  \bibinfo{author}{\bibfnamefont{H.}~\bibnamefont{Ritsch}}, \bibnamefont{and}
  \bibinfo{author}{\bibfnamefont{H.-C.} \bibnamefont{N\"agerl}},
  \bibinfo{journal}{Science} \textbf{\bibinfo{volume}{321}},
  \bibinfo{pages}{1062} (\bibinfo{year}{2008}).

\bibitem[{\citenamefont{Sage et~al.}(2005)\citenamefont{Sage, Sainis, Bergeman,
  and DeMille}}]{SagJer05}
\bibinfo{author}{\bibfnamefont{J.~M.} \bibnamefont{Sage}},
  \bibinfo{author}{\bibfnamefont{S.}~\bibnamefont{Sainis}},
  \bibinfo{author}{\bibfnamefont{T.}~\bibnamefont{Bergeman}}, \bibnamefont{and}
  \bibinfo{author}{\bibfnamefont{D.}~\bibnamefont{DeMille}},
  \bibinfo{journal}{Phys. Rev. Lett.} \textbf{\bibinfo{volume}{94}},
  \bibinfo{pages}{203001} (\bibinfo{year}{2005}).

\bibitem[{\citenamefont{Kerman et~al.}(2004)\citenamefont{Kerman, Sage, Sainis,
  Bergeman, and DeMille}}]{KerSag04}
\bibinfo{author}{\bibfnamefont{A.~J.} \bibnamefont{Kerman}},
  \bibinfo{author}{\bibfnamefont{J.~M.} \bibnamefont{Sage}},
  \bibinfo{author}{\bibfnamefont{S.}~\bibnamefont{Sainis}},
  \bibinfo{author}{\bibfnamefont{T.}~\bibnamefont{Bergeman}}, \bibnamefont{and}
  \bibinfo{author}{\bibfnamefont{D.}~\bibnamefont{DeMille}},
  \bibinfo{journal}{Phys. Rev. Lett.} \textbf{\bibinfo{volume}{92}},
  \bibinfo{pages}{033004} (\bibinfo{year}{2004}).

\bibitem[{\citenamefont{Wang et~al.}(2004)\citenamefont{Wang, Qi, Stone,
  Nikolayeva, Wang, Hattaway, Gensemer, Gould, Eyler, and Stwalley}}]{WanQi04}
\bibinfo{author}{\bibfnamefont{D.}~\bibnamefont{Wang}},
  \bibinfo{author}{\bibfnamefont{J.}~\bibnamefont{Qi}},
  \bibinfo{author}{\bibfnamefont{M.~F.} \bibnamefont{Stone}},
  \bibinfo{author}{\bibfnamefont{O.}~\bibnamefont{Nikolayeva}},
  \bibinfo{author}{\bibfnamefont{H.}~\bibnamefont{Wang}},
  \bibinfo{author}{\bibfnamefont{B.}~\bibnamefont{Hattaway}},
  \bibinfo{author}{\bibfnamefont{S.~D.} \bibnamefont{Gensemer}},
  \bibinfo{author}{\bibfnamefont{P.~L.} \bibnamefont{Gould}},
  \bibinfo{author}{\bibfnamefont{E.~E.} \bibnamefont{Eyler}}, \bibnamefont{and}
  \bibinfo{author}{\bibfnamefont{W.~C.} \bibnamefont{Stwalley}},
  \bibinfo{journal}{Phys. Rev. Lett.} \textbf{\bibinfo{volume}{93}},
  \bibinfo{pages}{243005} (\bibinfo{year}{2004}).

\bibitem[{\citenamefont{Haimberger et~al.}(2004)\citenamefont{Haimberger,
  Kleinert, Bhattacharya, and Bigelow}}]{HaiKle04}
\bibinfo{author}{\bibfnamefont{C.}~\bibnamefont{Haimberger}},
  \bibinfo{author}{\bibfnamefont{J.}~\bibnamefont{Kleinert}},
  \bibinfo{author}{\bibfnamefont{M.}~\bibnamefont{Bhattacharya}},
  \bibnamefont{and} \bibinfo{author}{\bibfnamefont{N.~P.}
  \bibnamefont{Bigelow}}, \bibinfo{journal}{Phys. Rev. A}
  \textbf{\bibinfo{volume}{70}}, \bibinfo{pages}{021402}
  (\bibinfo{year}{2004}).

\bibitem[{\citenamefont{Nemitz et~al.}(2009)\citenamefont{Nemitz, Baumer,
  M\"unchow, Tassy, and G\"orlitz}}]{NemBau09}
\bibinfo{author}{\bibfnamefont{N.}~\bibnamefont{Nemitz}},
  \bibinfo{author}{\bibfnamefont{F.}~\bibnamefont{Baumer}},
  \bibinfo{author}{\bibfnamefont{F.}~\bibnamefont{M\"unchow}},
  \bibinfo{author}{\bibfnamefont{S.}~\bibnamefont{Tassy}}, \bibnamefont{and}
  \bibinfo{author}{\bibfnamefont{A.}~\bibnamefont{G\"orlitz}},
  \bibinfo{journal}{Phys. Rev. A} \textbf{\bibinfo{volume}{79}},
  \bibinfo{pages}{061403} (\bibinfo{year}{2009}).

\bibitem[{\citenamefont{Wang and Stwalley}(1998)}]{WanStw98}
\bibinfo{author}{\bibfnamefont{H.}~\bibnamefont{Wang}} \bibnamefont{and}
  \bibinfo{author}{\bibfnamefont{W.~C.} \bibnamefont{Stwalley}},
  \bibinfo{journal}{J. Chem. Phys.} \textbf{\bibinfo{volume}{108}},
  \bibinfo{pages}{5767} (\bibinfo{year}{1998}).

\bibitem[{\citenamefont{Rousseau et~al.}(1999)\citenamefont{Rousseau, Allouche,
  Aubert-Fr\'econ, Magnier, Kowalczyk, and Jastrzebski}}]{RouAll99}
\bibinfo{author}{\bibfnamefont{S.}~\bibnamefont{Rousseau}},
  \bibinfo{author}{\bibfnamefont{A.~R.} \bibnamefont{Allouche}},
  \bibinfo{author}{\bibfnamefont{M.}~\bibnamefont{Aubert-Fr\'econ}},
  \bibinfo{author}{\bibfnamefont{S.}~\bibnamefont{Magnier}},
  \bibinfo{author}{\bibfnamefont{P.}~\bibnamefont{Kowalczyk}},
  \bibnamefont{and}
  \bibinfo{author}{\bibfnamefont{W.}~\bibnamefont{Jastrzebski}},
  \bibinfo{journal}{Chem. Phys.} \textbf{\bibinfo{volume}{247}},
  \bibinfo{pages}{193} (\bibinfo{year}{1999}).

\bibitem[{\citenamefont{Jastrzebski et~al.}(2001)\citenamefont{Jastrzebski,
  Kowalczyk, and Pashov}}]{JasKow01}
\bibinfo{author}{\bibfnamefont{W.}~\bibnamefont{Jastrzebski}},
  \bibinfo{author}{\bibfnamefont{P.}~\bibnamefont{Kowalczyk}},
  \bibnamefont{and} \bibinfo{author}{\bibfnamefont{A.}~\bibnamefont{Pashov}},
  \bibinfo{journal}{J. Mol. Spectr.} \textbf{\bibinfo{volume}{209}},
  \bibinfo{pages}{50} (\bibinfo{year}{2001}).

\bibitem[{\citenamefont{Salami et~al.}(2007)\citenamefont{Salami, Ross, Crozet,
  Jastrzebski, Kowalczyk, and LeRoy}}]{SalRos07}
\bibinfo{author}{\bibfnamefont{H.}~\bibnamefont{Salami}},
  \bibinfo{author}{\bibfnamefont{A.}~\bibnamefont{Ross}},
  \bibinfo{author}{\bibfnamefont{J.}~\bibnamefont{Crozet}},
  \bibinfo{author}{\bibfnamefont{W.}~\bibnamefont{Jastrzebski}},
  \bibinfo{author}{\bibfnamefont{P.}~\bibnamefont{Kowalczyk}},
  \bibnamefont{and} \bibinfo{author}{\bibfnamefont{R.~J.} \bibnamefont{LeRoy}},
  \bibinfo{journal}{J. Chem. Phys.} \textbf{\bibinfo{volume}{126}},
  \bibinfo{pages}{194313} (\bibinfo{year}{2007}).

\bibitem[{\citenamefont{Tiemann et~al.}(2009)\citenamefont{Tiemann, Kn\"ockel,
  Kowalczyk, Jastrzebski, Pashov, Salami, and Ross}}]{TieKno09}
\bibinfo{author}{\bibfnamefont{E.}~\bibnamefont{Tiemann}},
  \bibinfo{author}{\bibfnamefont{H.}~\bibnamefont{Kn\"ockel}},
  \bibinfo{author}{\bibfnamefont{P.}~\bibnamefont{Kowalczyk}},
  \bibinfo{author}{\bibfnamefont{W.}~\bibnamefont{Jastrzebski}},
  \bibinfo{author}{\bibfnamefont{A.}~\bibnamefont{Pashov}},
  \bibinfo{author}{\bibfnamefont{H.}~\bibnamefont{Salami}}, \bibnamefont{and}
  \bibinfo{author}{\bibfnamefont{A.~J.} \bibnamefont{Ross}},
  \bibinfo{journal}{Phys. Rev. A} \textbf{\bibinfo{volume}{79}},
  \bibinfo{pages}{042716} (\bibinfo{year}{2009}).

\bibitem[{\citenamefont{Bussery et~al.}(1987)\citenamefont{Bussery, Achkar, and
  Aubert-Fr\'econ}}]{BusAch87}
\bibinfo{author}{\bibfnamefont{B.}~\bibnamefont{Bussery}},
  \bibinfo{author}{\bibfnamefont{Y.}~\bibnamefont{Achkar}}, \bibnamefont{and}
  \bibinfo{author}{\bibfnamefont{M.}~\bibnamefont{Aubert-Fr\'econ}},
  \bibinfo{journal}{Chem. Phys.} \textbf{\bibinfo{volume}{116}},
  \bibinfo{pages}{319} (\bibinfo{year}{1987}).

\bibitem[{\citenamefont{Movre and Beuc}(1985)}]{MovBeu85}
\bibinfo{author}{\bibfnamefont{M.}~\bibnamefont{Movre}} \bibnamefont{and}
  \bibinfo{author}{\bibfnamefont{R.}~\bibnamefont{Beuc}},
  \bibinfo{journal}{Phys. Rev. A} \textbf{\bibinfo{volume}{31}},
  \bibinfo{pages}{2957} (\bibinfo{year}{1985}).

\bibitem[{\citenamefont{Krems et~al.}(2009)\citenamefont{Krems, Stwalley, and
  Bretislav}}]{RomKre09}
\bibinfo{editor}{\bibfnamefont{R.~V.} \bibnamefont{Krems}},
  \bibinfo{editor}{\bibfnamefont{W.~C.} \bibnamefont{Stwalley}},
  \bibnamefont{and}
  \bibinfo{editor}{\bibfnamefont{F.}~\bibnamefont{Bretislav}}, eds.,
  \emph{\bibinfo{title}{Cold Molecules: Theory, experiment, applications}}
  (\bibinfo{publisher}{CRC Press}, \bibinfo{address}{Boca Raton, Florida},
  \bibinfo{year}{2009}).

\bibitem[{\citenamefont{Ridinger et~al.}(2011)\citenamefont{Ridinger,
  Chaudhuri, Salez, Eismann, Fernandes, Magalh\~{a}es, Wilkowski, Salomon, and
  Chevy}}]{RidCha11}
\bibinfo{author}{\bibfnamefont{A.}~\bibnamefont{Ridinger}},
  \bibinfo{author}{\bibfnamefont{S.}~\bibnamefont{Chaudhuri}},
  \bibinfo{author}{\bibfnamefont{T.}~\bibnamefont{Salez}},
  \bibinfo{author}{\bibfnamefont{U.}~\bibnamefont{Eismann}},
  \bibinfo{author}{\bibfnamefont{D.~R.} \bibnamefont{Fernandes}},
  \bibinfo{author}{\bibfnamefont{K.}~\bibnamefont{Magalh\~{a}es}},
  \bibinfo{author}{\bibfnamefont{D.}~\bibnamefont{Wilkowski}},
  \bibinfo{author}{\bibfnamefont{C.}~\bibnamefont{Salomon}}, \bibnamefont{and}
  \bibinfo{author}{\bibfnamefont{F.}~\bibnamefont{Chevy}},
  \bibinfo{journal}{Eur. Phys. J. D}  (\bibinfo{year}{2011}).

\bibitem[{\citenamefont{F\"uhrer and Walther}(2008)}]{FueWal08}
\bibinfo{author}{\bibfnamefont{T.}~\bibnamefont{F\"uhrer}} \bibnamefont{and}
  \bibinfo{author}{\bibfnamefont{T.}~\bibnamefont{Walther}},
  \bibinfo{journal}{Opt. Lett.} \textbf{\bibinfo{volume}{33}},
  \bibinfo{pages}{372} (\bibinfo{year}{2008}).

\bibitem[{\citenamefont{Ridinger}(2011)}]{Rid11}
\bibinfo{author}{\bibfnamefont{A.}~\bibnamefont{Ridinger}},
  \emph{\bibinfo{title}{Ph.D.~thesis}} (\bibinfo{publisher}{\'Ecole Normale
  Sup\'erieure}, \bibinfo{address}{Paris}, \bibinfo{year}{2011}).

\bibitem[{\citenamefont{LeRoy and Bernstein}(1970)}]{LeRBer70}
\bibinfo{author}{\bibfnamefont{J.~R.} \bibnamefont{LeRoy}} \bibnamefont{and}
  \bibinfo{author}{\bibfnamefont{R.~B.} \bibnamefont{Bernstein}},
  \bibinfo{journal}{J. Chem. Phys.} \textbf{\bibinfo{volume}{52}},
  \bibinfo{pages}{3869} (\bibinfo{year}{1970}).

\bibitem[{\citenamefont{Stwalley}(1970)}]{Stw70}
\bibinfo{author}{\bibfnamefont{W.~C.} \bibnamefont{Stwalley}},
  \bibinfo{journal}{Chem. Phys. Lett.} \textbf{\bibinfo{volume}{6}},
  \bibinfo{pages}{241} (\bibinfo{year}{1970}).

\bibitem[{\citenamefont{Williams et~al.}(1996)\citenamefont{Williams, Tiesinga,
  and Julienne}}]{WilTie96}
\bibinfo{author}{\bibfnamefont{C.~J.} \bibnamefont{Williams}},
  \bibinfo{author}{\bibfnamefont{E.}~\bibnamefont{Tiesinga}}, \bibnamefont{and}
  \bibinfo{author}{\bibfnamefont{P.~S.} \bibnamefont{Julienne}},
  \bibinfo{journal}{Phys. Rev. A} \textbf{\bibinfo{volume}{53}},
  \bibinfo{pages}{R1939} (\bibinfo{year}{1996}).

\bibitem[{\citenamefont{Grochola et~al.}(2009)\citenamefont{Grochola, Pashov,
  Deiglmayr, Repp, Tiemann, Wester, and Weidem\"uller}}]{GroPas09}
\bibinfo{author}{\bibfnamefont{A.}~\bibnamefont{Grochola}},
  \bibinfo{author}{\bibfnamefont{A.}~\bibnamefont{Pashov}},
  \bibinfo{author}{\bibfnamefont{J.}~\bibnamefont{Deiglmayr}},
  \bibinfo{author}{\bibfnamefont{M.}~\bibnamefont{Repp}},
  \bibinfo{author}{\bibfnamefont{E.}~\bibnamefont{Tiemann}},
  \bibinfo{author}{\bibfnamefont{R.}~\bibnamefont{Wester}}, \bibnamefont{and}
  \bibinfo{author}{\bibfnamefont{M.}~\bibnamefont{Weidem\"uller}},
  \bibinfo{journal}{J. Chem. Phys.} \textbf{\bibinfo{volume}{131}},
  \bibinfo{pages}{054304} (\bibinfo{year}{2009}).

\bibitem[{\citenamefont{Azizi et~al.}(2004)\citenamefont{Azizi, Aymar, and
  Dulieu}}]{AziAym04}
\bibinfo{author}{\bibfnamefont{S.}~\bibnamefont{Azizi}},
  \bibinfo{author}{\bibfnamefont{M.}~\bibnamefont{Aymar}}, \bibnamefont{and}
  \bibinfo{author}{\bibfnamefont{O.}~\bibnamefont{Dulieu}},
  \bibinfo{journal}{Eur. Phys. J. D} \textbf{\bibinfo{volume}{31}},
  \bibinfo{pages}{195} (\bibinfo{year}{2004}).

\bibitem[{\citenamefont{Pillet et~al.}(1997)\citenamefont{Pillet, Crubellier,
  Bleton, Dulieu, Nosbaum, Mourachko, and Masnou-Seeuws}}]{PilCru97}
\bibinfo{author}{\bibfnamefont{P.}~\bibnamefont{Pillet}},
  \bibinfo{author}{\bibfnamefont{A.}~\bibnamefont{Crubellier}},
  \bibinfo{author}{\bibfnamefont{A.}~\bibnamefont{Bleton}},
  \bibinfo{author}{\bibfnamefont{O.}~\bibnamefont{Dulieu}},
  \bibinfo{author}{\bibfnamefont{P.}~\bibnamefont{Nosbaum}},
  \bibinfo{author}{\bibfnamefont{I.}~\bibnamefont{Mourachko}},
  \bibnamefont{and}
  \bibinfo{author}{\bibfnamefont{F.}~\bibnamefont{Masnou-Seeuws}},
  \bibinfo{journal}{J. Phys. B} \textbf{\bibinfo{volume}{30}},
  \bibinfo{pages}{2801} (\bibinfo{year}{1997}).

\bibitem[{\citenamefont{Pashov et~al.}(1998)\citenamefont{Pashov, Jastrzebski,
  and Kowalczyk}}]{PasJas98}
\bibinfo{author}{\bibfnamefont{A.}~\bibnamefont{Pashov}},
  \bibinfo{author}{\bibfnamefont{W.}~\bibnamefont{Jastrzebski}},
  \bibnamefont{and}
  \bibinfo{author}{\bibfnamefont{P.}~\bibnamefont{Kowalczyk}},
  \bibinfo{journal}{Chem. Phys. Lett.} \textbf{\bibinfo{volume}{292}},
  \bibinfo{pages}{615} (\bibinfo{year}{1998}).

\end{thebibliography}
%

\end{document}